\documentclass[a4paper,11pt]{article}
\usepackage{pos}
\usepackage{booktabs}

\title{An AGN-starburst composite multi-messenger model of NGC 1068}
 \ShortTitle{An AGN-starburst composite multi-messenger model of NGC 1068}

\author*[a]{Bj\"{o}rn Eichmann}
\author[b]{Ralf-Jürgen Dettmar}
\author[a]{Julia Becker Tjus}

\affiliation[a]{RAPP Center, Ruhr-Universit\"at Bochum, Institut f\"ur Theoretische Physik IV \\ 44780 Bochum, Germany}
\affiliation[b]{RAPP Center, Ruhr-Universit\"at Bochum, Astronomical Institute (AIRUB) \\ 44780 Bochum, Germany}

\emailAdd{eiche@tp4.rub.de}
\emailAdd{dettmar@astro.rub.de}
\emailAdd{julia.tjus@rub.de}

\abstract{Recent multi-wavelength observations indicate that some starburst galaxies show a dominant nonthermal contribution from its central region. These active galactic nuclei (AGN)-starburst composites are of special interest, as both phenomena on their own are potential sources of the high-energetic cosmic rays and their accompanied gamma-ray and neutrino emission. Here, we will focus on NGC 1068, which is known since several years from its atypical radio-gamma-ray correlation. Recently this source has also shown strong indications of high energy neutrino emission. A first semi-analytical, two-component multi-messenger model is presented that already gives some constraints on the AGN-starburst composite characteristics of NGC 1068 and exposes the need to include both starburst \emph{and} AGN corona to describe the multi-messenger data.}

\FullConference{37$^{\rm{th}}$ International Cosmic Ray Conference (ICRC 2021)\\
		July 12th -- 23rd, 2021\\
		Online -- Berlin, Germany}

\begin{document}
\maketitle

\section{Introduction}
Starburst galaxies, that are characterized by a high star-formation rate leading to a large number of cosmic-ray-accelerating supernova remnants, and the accretion disk-fed black hole regions in galactic nuclei, so-called active galactic nuclei (AGN), are both phenomena on their own that generate high-energetic cosmic rays (HECRs).
Associated to these cosmic-rays, multiwavelength observations have shown that AGN and starburst galaxies provide non-thermal emission over a broad range of energies. With first Fermi detections \citep{a5:fermi_starbursts2012}, starburst galaxies started to become visible in high-energy gamma rays. Moreover, some starburst galaxies, such as NGC\,1068, NGC\,4945 or Arp\,220, show a dominant non-thermal contribution from their central regions, indicating the presence of an active black hole. These active galactic nuclei (AGN)-starburst composites are of special interest, as the simultaneous energy release by multiple supernova events as well as by accretion and possible jet activity of the central black hole makes AGN-starburst composites very promising candidates for the direct detection of HECR and high-energy astrophysical neutrino sources. The latter is in particular the case for NGC\,1068, since a recent search by the IceCube experiment for sources of high-energy neutrino emission \citep{a5:IceCube2020} exposed the direction of NGC\,1068 as the most significant one in the sky, with $2.9\upsigma$. With respect to the non-thermal emission of photons, neutrinos, and HECRs from these astrophysical objects, the AGN and the surrounding starburst are usually discussed individually, e.g.~\cite{a5:eichmann2016,a5:Yoast-Hull-etal-2014,a5:Murase+2020}, leaving many open questions on the leading interaction processes within the AGN corona compared to the surrounding starburst, on their interplay, and on the impact of the torus that is typically in between these environments. 
The observed large-scale motions, in particular the strong ionized gas outflow in the inner hundreds of parsec or the even larger galactic superwinds as, e.g., observed in NGC\,3079 \citep{a5:Hodges-Kluck_2020},
indicate that the starburst and the AGN activity are connected, though the particle and photon target fields of high-energy CRs change significantly. For instance, continuous and line emission at infrared wavelength dominate at the torus and the starburst region \cite{a5:eichmann2016}, whereas the optical/UV emission by the accretion disk that gets comptonized to X-ray energies becomes important within the AGN corona \cite{a5:Murase+2020}.

\section{The AGN-starburst model}
In the long term perspective, we need to correlate the circumnuclear starburst region with the inner AGN activity including the impact by the intermediate torus region, to provide a complete description of the non-thermal phenomena of AGN-starburst composite galaxies. However, at first we perform a baseline study where we elaborate the different physical processes by a semi-analytical two-zone model --- the starburst zone and the AGN corona zone. Hereby, we suppose that in both zones diffusive shock acceleration yields a differential source rate $q(E)$ of relativistic protons and electrons that can be described by a power-law distribution up to a certain maximal kinetic energy $E_{\rm max}$. In the case of the starburst zone, we suppose that a certain fraction $f_{\rm SN}$ of the total energy of the supernova --- that occurs with an approximate rate \cite{Veilleux+2005} $\nu_{\rm SN}\simeq0.34\,[L_{\rm IR}/(10^{11}\,L_\odot)]\text{yr}^{-1}$ dependent on the IR luminosity $L_{\rm IR}$ --- gets accelerated to relativistic energies. For the AGN corona, we suppose that a fraction $f_{\rm inj}\ll 1$ of the X-ray emissivity at an injection energy of $E_{\rm inj}$ goes into relativistic protons. For both zones, we apply that the nonthermal electrons are normalized by the nonthermal proton rates due to the requested quasi-neutral total charge number above kinetic energies of $E_{\rm inj}$ \cite{a5:eichmann2016}. 

After the injection of nonthermal protons and electrons into these zones, they suffer from continuous energy losses or escape from these zones. So, the steady-state behavior of the differential electron and proton density $n(E)$ can be approximated by
\begin{equation}
-\frac{\partial}{\partial E}\left(\frac{E\,n(E)}{\tau_{\mathsf{cool}}(E)}\right)=q(E)-\frac{n(E)}{\tau_{\mathsf{esc}}(E)}\,.
\label{eq:teq0}
\end{equation}
Here, $\tau_{\mathsf{cool}}$ refers to the total continuous energy-loss timescale, which in case of the relativistic electrons is given by the inverse of the sum of the synchrotron, inverse Compton (IC), non-thermal Bremsstrahlung, and Coulomb loss rates, according to
\begin{equation}
    \tau_{\mathsf{cool}}^{\rm (e)}=[\tau_{\mathsf{syn}}^{-1}+\tau_{\mathsf{ic}}^{-1}+\tau_{\mathsf{brems}}^{-1}+\tau_{\mathsf{C}}^{-1}]^{-1}\,,
\end{equation}
in case of the relativistic protons 
\begin{equation}
    \tau_{\mathsf{cool}}^{\rm (p)}=[\tau_{\mathsf{syn}}^{-1}+(\tau_{\mathsf{p\gamma}}^\pi)^{-1}+(\tau_{\mathsf{p\gamma}}^{\rm ee})^{-1}+\tau_{\mathsf{pp}}^{-1}]^{-1}\,,
\end{equation}
including the photopion ($\pi$), photopair (ee), and hadronic pion (pp) production loss rates. For the latter interaction we account for the spectral hardening at a few hundreds of GeV \cite{Krakau+2015}, and all details on the other individual loss rates can be found in monographs, e.g.~\cite{Schlickeiser2002_book, Dermer+2009_book}. The total escape rate via gyro-resonant scattering through turbulence with a power-law spectrum $\propto k^{-\varkappa}$ and a turbulence strength $\eta^{-1}$ as well as a bulk stream flow can be approximated by \cite{a5:Murase+2020}
\begin{equation}
    \tau_{\mathsf{esc}}\simeq\begin{cases}
    \left[ \frac{\eta}{9}\,\frac{c}{R}\left( \frac{e\,B\,R}{E} \right)^{\varkappa-2} + \frac{v_{\rm w}}{R} \right]^{-1}\,,\quad&\text{for the starburst}\\
    \left[ \frac{\eta}{9}\,\frac{c}{R}\left( \frac{e\,B\,R}{E} \right)^{\varkappa-2} + \frac{\alpha v_{\rm K}}{R} \right]^{-1}\,,\quad&\text{for the AGN corona}\,,
    \end{cases}
\end{equation}
with respect to the characteristic size $R$ and the magnetic field strength of the zones. In the case of the starburst zone, the bulk motion is given by a galactic wind with a velocity $v_{\rm w}$; and in the case of the AGN corona, we account for the infall timescale, which is expected to be similar to the advection dominated accretion flow \cite{a5:Murase+2020}, with a viscosity parameter $\alpha$ and the Keplerian velocity $v_{\rm K}=\sqrt{G\,M_{\rm BH}/R}$. Note that in the case of the relativistic electrons $q(E)$ includes also the secondary electrons that emerge from the photomeson and the hadronic pion production. We are aware that especially in the AGN corona rather stochastic diffuse acceleration (SDA) than diffusive shock acceleration (DSA) is happening, but also first-order Fermi acceleration or magnetic reconnection might accelerate non-thermal particles very efficiently \cite{Kheirandish+2021_arXiv}, so that, for this baseline study, we keep a simple power-law source spectrum and include no further details on the acceleration processes in the transport equation \ref{eq:teq0}. 

\begin{table}[h!]
\centering
\caption{Parameters of NGC 1068.}
Starburst \\
  \begin{tabular}{c c c c c c c c c c c c} 
  \toprule
              $\gamma$ & $E_{\rm inj}$ & $f_{\rm SN}$ & $L_{\rm IR}$ & $E_{\rm max}^{\rm (p)}$ & $E_{\rm max}^{\rm (e)}$ & $n_{\rm th}^{\rm (p)}$ & $B$ & $\eta$ & $\varkappa$  & $R$ & $v_{\rm w}$  \\ 
              & [MeV] & & $[10^{11}L_\odot]$ & [TeV] & [TeV] & $[\text{cm}^{-3}]$ & [mG] & & & [pc] & [km/s]  \\
   \midrule
    $2$ & $0.01$ & $0.06$ & $1$ & $100$ & $0.1$ & $100$ & $0.6$ & $66.7$ & $5/3$ & $200$ & $100$  \\ 
   \bottomrule
\end{tabular} \\ \vspace{0.5cm}
AGN corona \\
  \begin{tabular}{c c c c c c c c c c c c} 
  \toprule
              $\gamma$ & $E_{\rm inj}$ & $f_{\rm inj}$ & $L_{\rm X}$ & $E_{\rm max}^{\rm (p)}$ & $E_{\rm max}^{\rm (e)}$ & $\tau_{\rm T}$ & $\beta$ & $\eta$ & $\varkappa$  & $R$ & $\alpha$  \\ 
              & [MeV] & & $[10^{11}L_\odot]$ & [TeV] & [MeV] & & & & & $[R_{\rm S}]$ &  \\
   \midrule
    $2$ & $0.01$ & $0.0003$ & $0.18$ & $100$ & $1$ & $0.5$ & $0.1$ & $10.$ & $5/3$ & $10$ & $0.1$  \\ 
   \bottomrule
\end{tabular}
  \label{tab:Param}
\end{table}

\begin{figure}[htbp]
    \centering
    \includegraphics[width=0.56\linewidth]{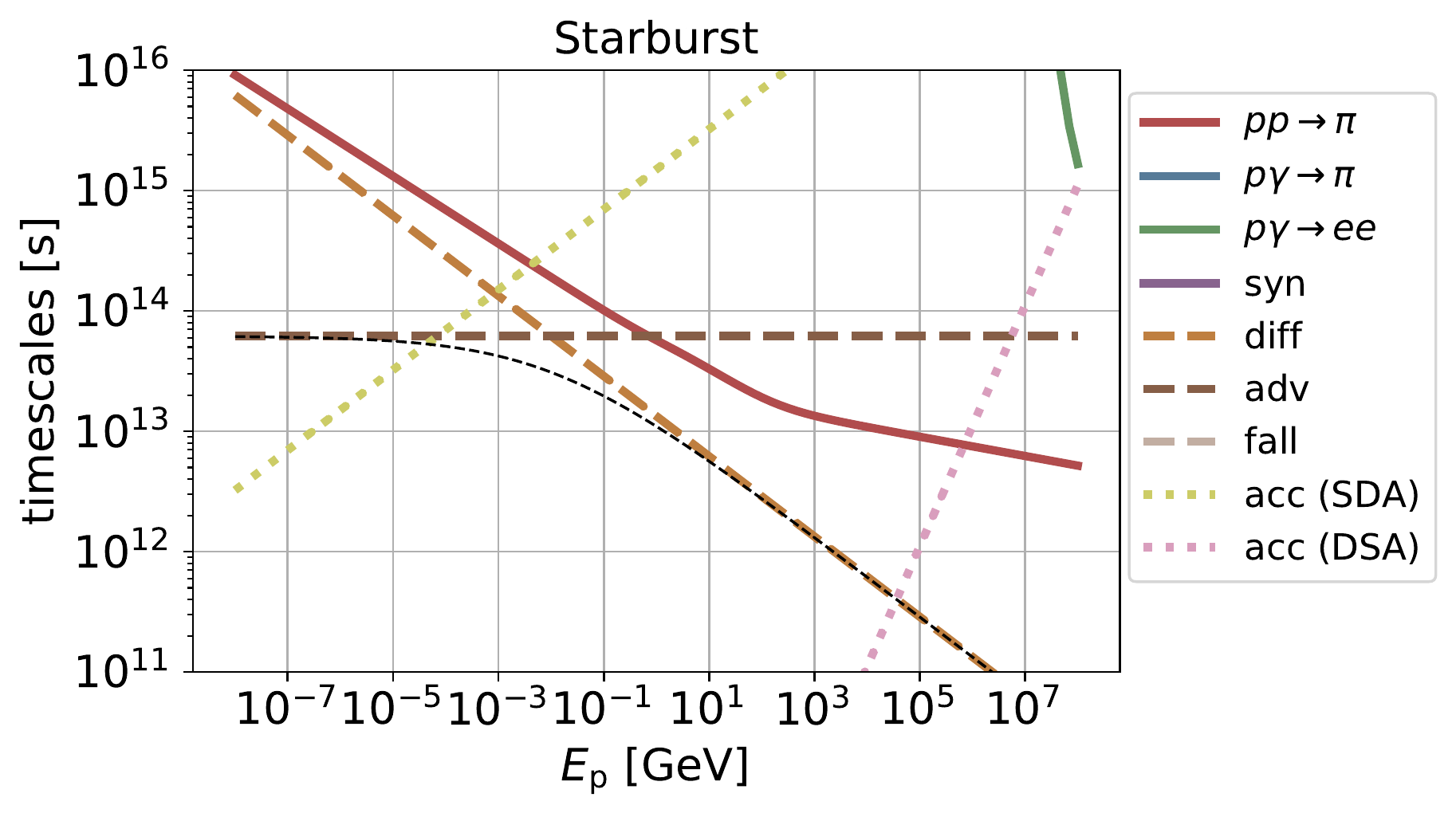}
    \includegraphics[width=0.43\linewidth]{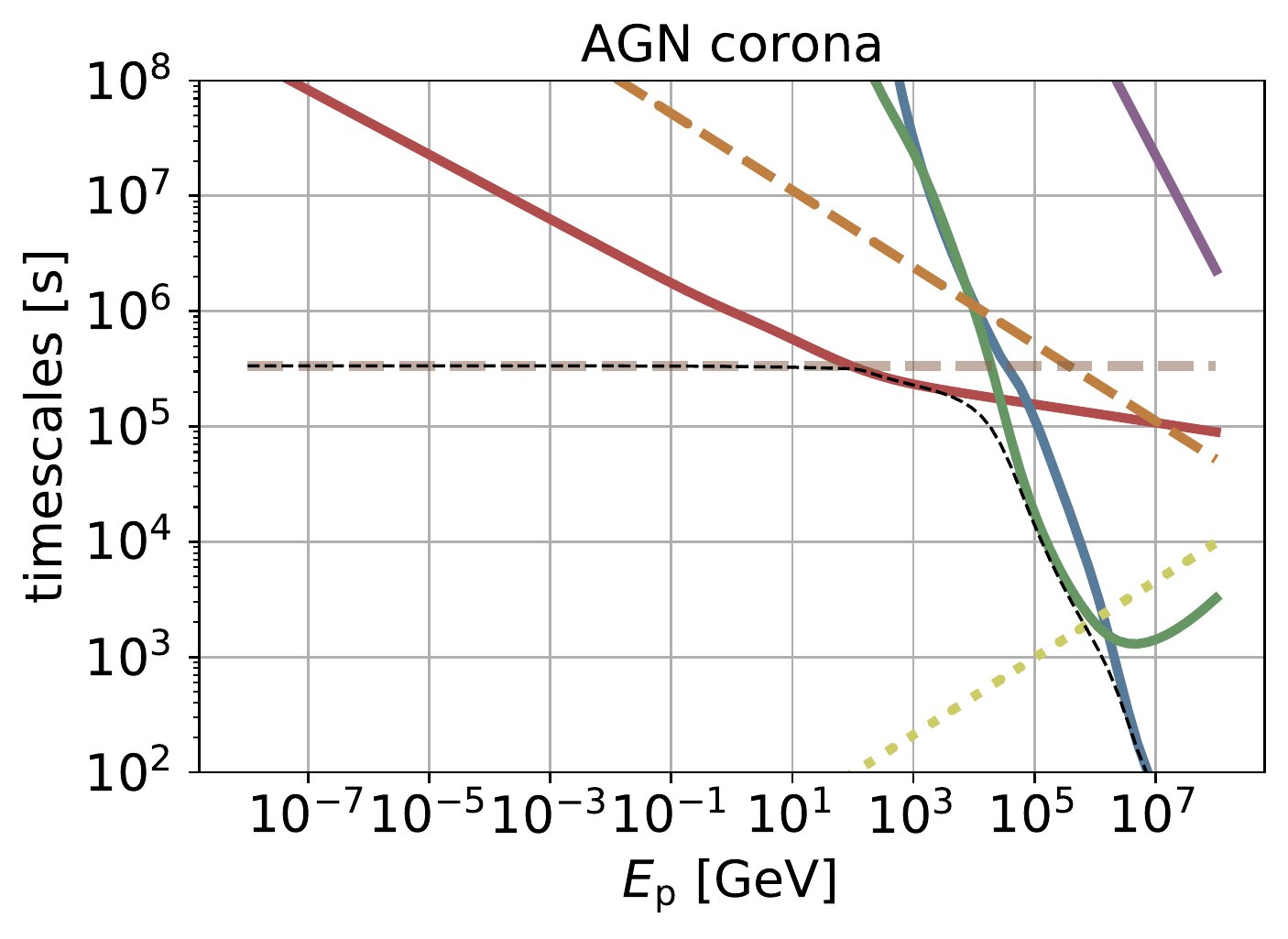}
    \includegraphics[width=0.56\linewidth]{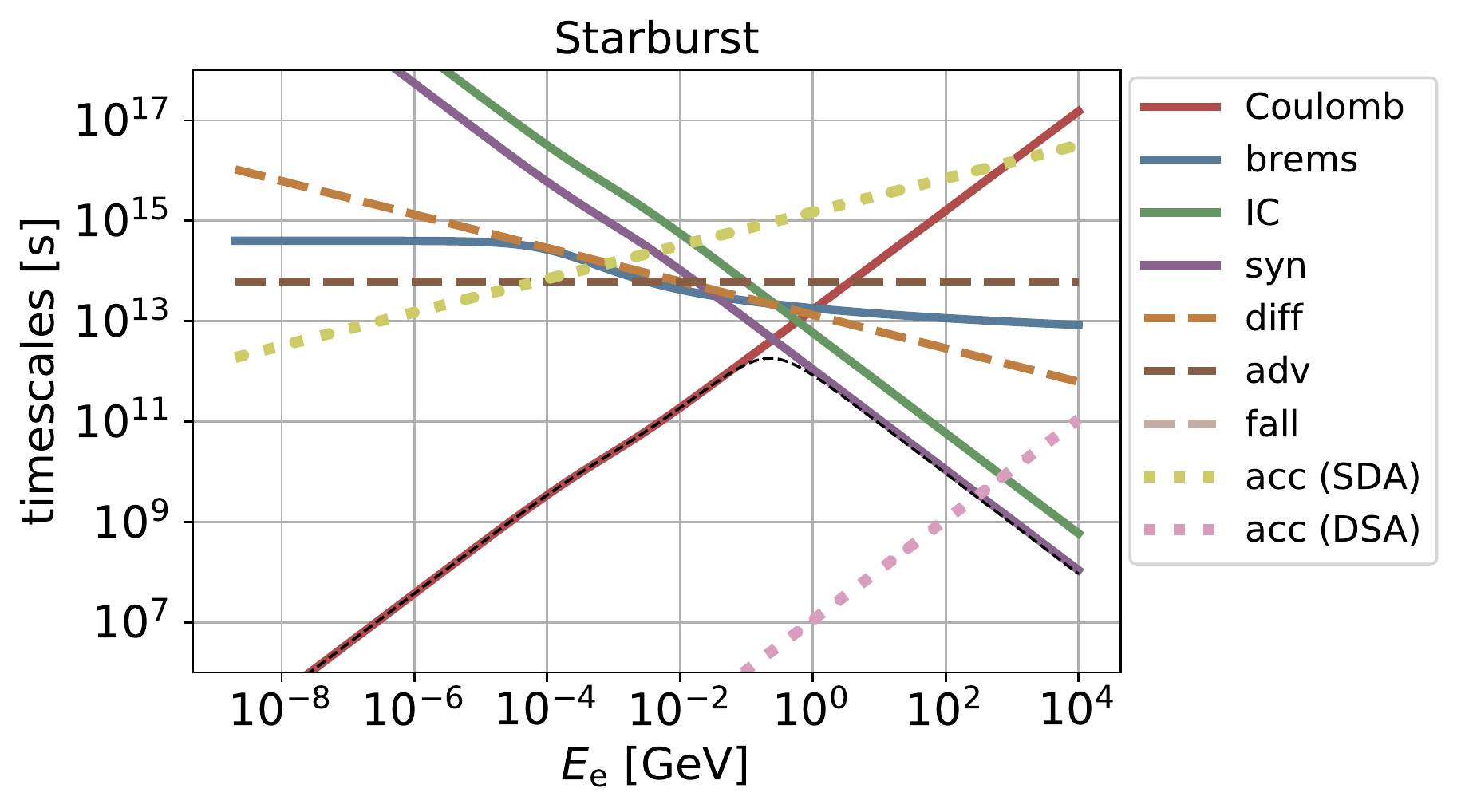}
    \includegraphics[width=0.43\linewidth]{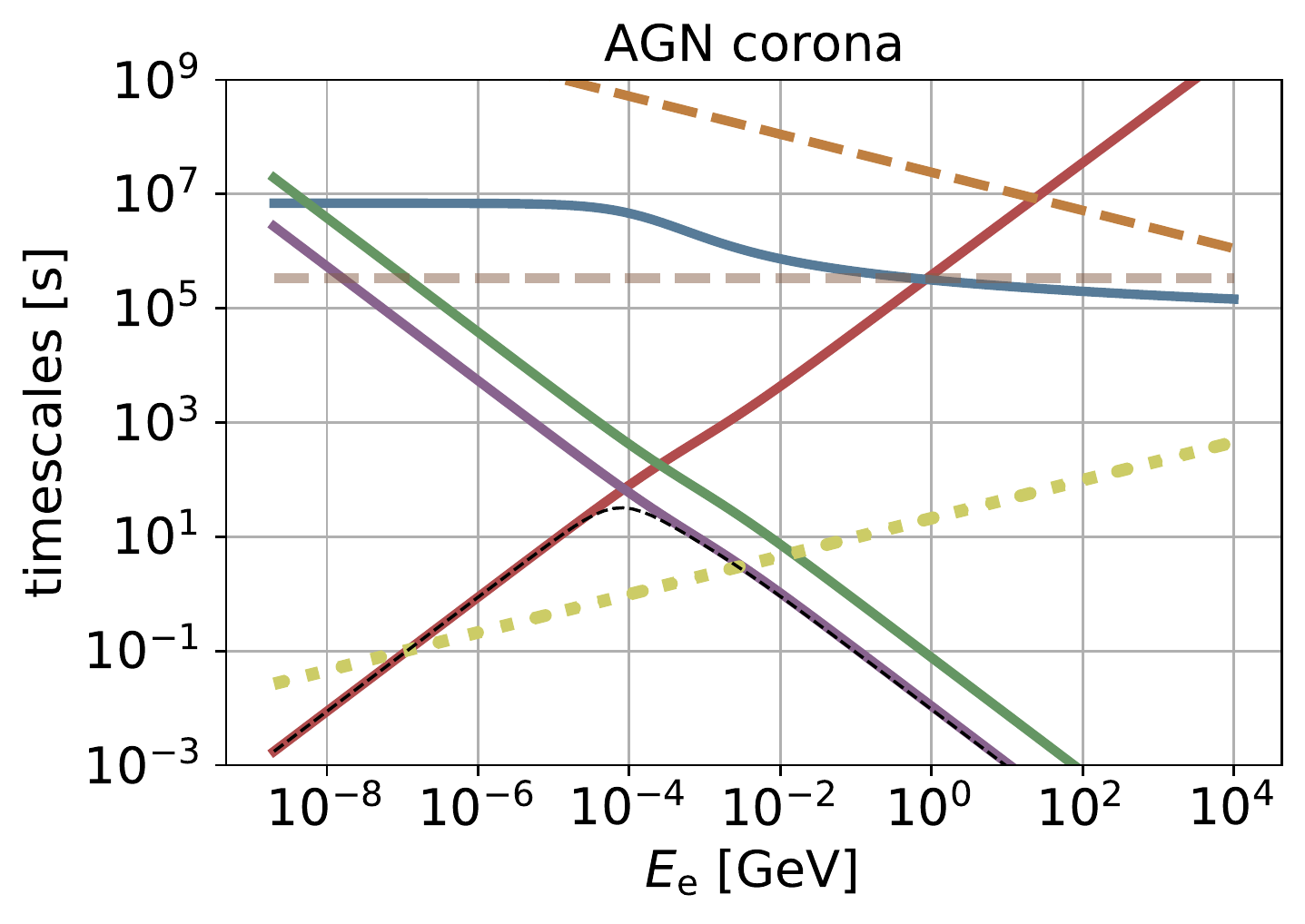}
    \caption{The different timescales of relativistic protons (\emph{upper panel}) and electrons (\emph{lower panel}) for the starburst zone (\emph{left}) and the AGN corona zone (\emph{left}). The supposed parameters are given in Table \ref{tab:Param}.}
    \label{fig:timescales} 
\end{figure}

The physical parameters, such as the magnetic field strength or the target densities of the thermal photon and gas distributions, are completely different in the starburst and the AGN zone, so that the resulting timescales $\tau_{\mathsf{cool}}$ and $\tau_{\mathsf{esc}}$ differ significantly (see Fig.~\ref{fig:timescales}) yielding different relativistic particle distributions in these zones. Based on the recent "hidden core" model of the AGN corona \cite{a5:Murase+2020}, we use the same parameters and describe the corona zone by a characteristic radius of the disk $R=f_{\rm R}\,R_{\rm S}$, given as a multitude $f_{\rm R}\gg 1$ of the Schwarzschild radius $R_{\rm S}$, and a height $H\simeq R/\sqrt{3}$, so that the target density can by determined according to $n_{\rm th}^{\rm (p)}\simeq\tau_T/(\sigma_T H)$ with the Thomson optical depth $\tau_{\rm T}$. The magnetic field strength of the corona can be derived from pressure equilibrium according to $B=\sqrt{8\pi\,n_{\rm th}^{\rm (p)}\,k_{\rm B}T_{\rm p}/\beta}$ using the virial temperature $T_{\rm p}=m_pc^2/(6f_{\rm R}k_{\rm B})$ and a plasma beta ($\beta$). 

The Fig.~\ref{fig:timescales} exposes that --- for the supposed parameters of NGC 1068 (see Table \ref{tab:Param}) --- protons become accelerated up to energies of about $E_{\rm max}^{\rm (p)}\sim (10^4\dots 10^5)\,\text{GeV}$ by DSA in case of the starburst and by SDA in case of the AGN corona, so that
\begin{equation}
    q^{\rm (p)}(E)=q_{0}^{\rm (p)}(E/E_{\rm inj})^{-\gamma}\,\Theta[E_{\rm max}^{\rm (p)}-E]
\end{equation} 
Further, it shows that the primary electrons of the AGN corona can hardly be accelerated to higher energies without suffering from Coulomb losses. Hence, we need to account for the high-energy secondary electrons $q_{\rm e}^{(2)}(E)$ that are injected via hadronic processes, such as the photo-pion or hadronic pion production, and include this additional source term, so that
\begin{equation}
    q^{\rm (e)}(E)= q_{0}^{\rm (e)}(E/E_{\rm inj})^{-\gamma}\,\Theta[E_{\rm max}^{\rm (e)}-E]+q^{\rm (e)}_{(2)}(E)\,.
\end{equation}
In doing so, we need to account for the different values of $E_{\rm max}^{\rm (e)}$ with respect to the considered zone: In case of the starburst DSA can provide a maximal primary electron energy of several hundreds of GeV, whereas in the AGN corona the electrons suffer from significant synchrotron/ IC losses at energies above a few MeV. In addition, the electrons have to overcome the Coulomb losses at low energies, and hence, need to be injected into the SDA process at about few keV.\footnote{Note, that we do not account for the possible steepening of the turbulent power spectrum at energies below the thermal proton energy, which would lengthen the acceleration time considerably.} So, if primary electrons can be accelerated via turbulence or magnetic reconnection within the AGN corona, they can hardly reach energies significantly above their rest mass. The issues that emerge from a missing primary electron contribution from the AGN corona will be discussed in some more detail in the following section. 

Further, the timescale plots indicate that --- for the supposed parameters --- the starburst zone of NGC\,1068 is not a proton calorimeter, i.e.\ a significant amount is expected to leave the starburst region by diffusive transport (above ~10\,MeV) before loosing its energy via hadronic pion production. In case of the AGN corona, only low-energy protons with a kinetic energy $\lesssim 100\,\text{GeV}$ leave this zone by falling into the black hole before loosing its energy. 

%

\section{Multi-messenger SED of NGC 1068}
\begin{figure}[htbp]
    \centering
    \includegraphics[width=0.99\linewidth]{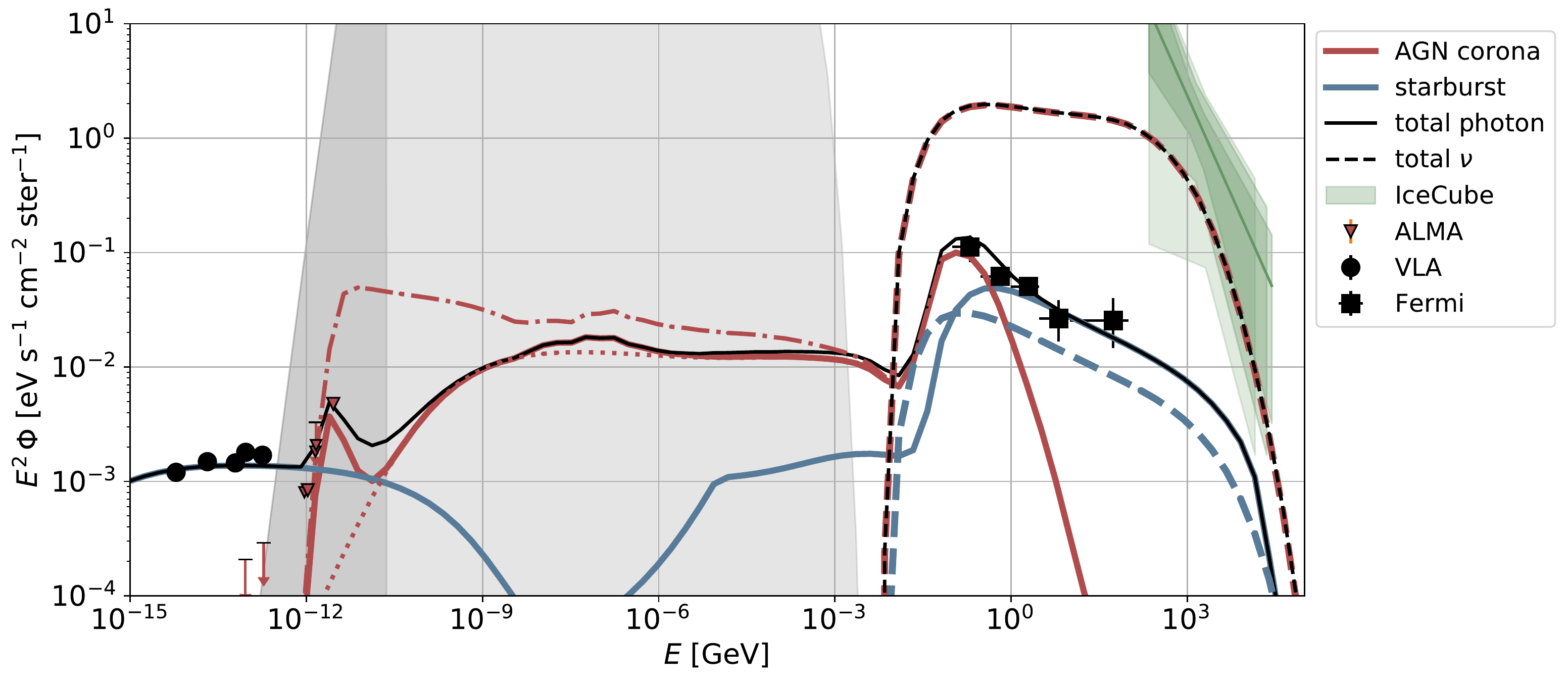}
    \caption{The model predictions of the \emph{photon (solid/ dotted/ dash-dotted lines)} and \emph{neutrino (dashed lines)} SED of NGC\,1068 with respect to the data (small red markers and upper limits refer to a beam size of $\sim 10\,\text{mas}$, and large black markers indicate a beam size of $\gtrsim 10\,\text{as}$). The light grey area indicates the energy range that is covered by the thermal photon fields (disk- and torus emission as well as Comptonized X-rays of the AGN corona) of the AGN and the dark grey area indicates the thermal IR emission of the starburst region. The red dotted line indicates the total photon SED by the AGN corona in the case of vanishing primary electron (or at least $E_{\rm max}^{\rm (e)}\ll 1\,\text{MeV}$) and the red dash-dotted line indicates the total photon SED by the AGN corona in the case of $E_{\rm max}^{\rm (e)}= 100\,\text{MeV}$.}
    \label{fig:SED} 
\end{figure}

Based on the differential proton and electron density, that results from the transport equation (\ref{eq:teq0}), we determine the nonthermal leptonic (synchrotron, inverse Compton and Bremsstrahlung) and hadronic (photo-pion and hadronic pion production) emission of these particles. In doing so we also need to account for the optical depth of the AGN corona, since this region gets optically thick in the radio at $E\lesssim 1\,\text{meV}$, due to synchrotron-self and free-free absorption, and in the gamma-ray regime at $E\gtrsim 1\,\text{GeV}$, due to $\gamma\gamma$-pair attenuation. Therefore, the AGN can only contribute at sub-GeV energies to the observed Fermi-flux and the starburst region takes over at higher energies (see Fig.~\ref{fig:SED}). In contrast to the gamma-rays, the high energy neutrinos that result from the hadronic pion production with the dense background plasma ($\sim (10^9\dots 10^{10})\,\text{cm}^{-3}$) can escape this central corona region and provide a good agreement with the recent IceCube measurements \cite{a5:IceCube2020}. 

Examining the Fig.~\ref{fig:SED} at its low energy regime, we assert that the IR data from ALMA \cite{Inoue+2020} can only be described by an emerging contribution by the AGN corona at about $1\,\text{meV}$, which is almost exactly the energy regime where the synchrotron emission of the corona becomes optically thin. Note that the ALMA data (as well as the upper limits) originate from the inner parsecs of NGC 1068 which excludes the starburst contribution. Using \emph{only the secondary electrons} generated by the hadronic interactions, the red dotted line indicates that we lack low energetic electrons (with energies below the pion rest mass), so that the ALMA and the sub-GeV Fermi data can hardly be explained at the same time. Adopting an efficient (pre-/re-)acceleration process for the primary electrons in the AGN corona so that $E_{\rm max}^{\rm (e)}\gg 1\,\text{MeV}$, the red dash-dotted line exposes that dip at some tens of meV vanishes. Hence, future observation of this source in the IR might clarify on the relevance of primary and secondary synchrotron radiation and the leptonic acceleration processes. 

\section{Conclusions}
Based on a simplified two-zone model of the starburst and the AGN corona of NGC\,1068 we could point out the need to the characteristic nonthermal emission processes of both regions to explain its broad-band spectral energy distribution. Hereby, we show that already this simplified model --- using reasonable parameters without any parameter fine-tuning --- provides not only a good agreement to the photon data, but also to the observed neutrino flux by IceCube. Here, our model indicates that the steep spectral behavior of the observed neutrino flux could actually result from the (exponential) cut-off of the primary CR protons in the AGN corona. Since not only the hadronic interaction processes but also the $\gamma\gamma$-pair production at GeV - TeV energies yield a multitude of electron-positron pairs in the AGN corona, these pairs likely cascade down to MeV energies, providing an additional MeV gamma-ray tail (see \cite{a5:Murase+2020}), that we have not taken into account so far. Either way, these pairs can hardly compensate for a possible lack of primary electrons at sub-MeV energies in case that Coulomb losses prohibit an efficient acceleration of thermal electrons. Further, we did not account for the faint radio jet structure \cite{Gallimore+1996,Gallimore+2004,Cotton+2008}, as its emission seems to be subdominant compared to the free-free emission \cite{Gallimore+2004,Cotton+2008}. However, this collimated plasma beam might provide an additional (pre-)accelerator for the primary electrons in the AGN corona. Including more details --- with respect to the physical processes\,/\,dependencies as well as the observational constraints --- and an extensive parameter study on both emission regions, we intend to provide an extended description of the whole multi-messenger SED of NGC\,1068 in the near future.


\bibliographystyle{JHEP}
\addcontentsline{toc}{section}{Bibliography}
\bibliography{references}
%

%
%
%

\end{document}